\documentclass[final,3p,times]{elsarticle}
\usepackage{amssymb,amsmath,amsthm}
\usepackage{lineno,natbib}
\usepackage{xcolor,soul}
\usepackage[colorlinks]{hyperref}
\usepackage{enumitem}
\usepackage{algorithm}
\usepackage{algpseudocode}
\usepackage{standalone}
\usepackage{booktabs}
\usepackage[OT1]{fontenc} 
\newtheorem{theorem}{Theorem}[section]

\newtheorem{remark}[theorem]{Remark}

\graphicspath{ {./Images/} }
\usepackage{graphicx}
\usepackage{float}
\usepackage{tikz}
\usepackage{caption}
\usepackage{subcaption}
\usepackage{multirow}
\usepackage{makecell}

\begin{document}
	\makeatletter
	\def\ps@pprintTitle{%
		\let\@oddhead\@empty
		\let\@evenhead\@empty
		\let\@oddfoot\@empty
		\let\@evenfoot\@oddfoot}
	\makeatother
	\begin{frontmatter}
		\title{Performance of Machine Learning Methods for Gravity Inversion: Successes and Challenges}
		
		\author[main]{Vahid Negahdari\corref{cor}}
		\ead{v.negahdari94@sharif.edu}

		\author[main]{Shirin Samadi Bahrami}
		\ead{shirin.samadibahrami@alum.sharif.edu}
		
		\author[main]{Seyed Reza Moghadasi}
		\ead{moghadasi@sharif.edu}
		
		\author[main]{Mohammad Reza Razvan}
		\ead{razvan@sharif.edu}

		\cortext[cor]{Corresponding author}	
		\address[main]{Department of Mathematical Sciences, Sharif University of Technology, Tehran, 11155-9415, Iran}

		\begin{abstract}
			
			Gravity inversion is the problem of estimating subsurface density distributions from observed gravitational field data. We consider the two-dimensional (2D) case, in which recovering density models from one-dimensional (1D) measurements leads to an underdetermined system with substantially more model parameters than measurements, making the inversion ill-posed and non-unique. Recent advances in machine learning have motivated data-driven approaches for gravity inversion. We first design a convolutional neural network (CNN) trained to directly map gravity anomalies to density fields, where a customized data structure is introduced to enhance the inversion performance. To further investigate generative modeling, we employ Variational Autoencoders (VAEs) and Generative Adversarial Networks (GANs), reformulating inversion as a latent-space optimization constrained by the forward operator. In addition, we assess whether classical iterative solvers such as Gradient Descent (GD), GMRES, LGMRES, and a recently proposed Improved Conjugate Gradient (ICG) method can refine CNN-based initial guesses and improve inversion accuracy. Our results demonstrate that CNN inversion not only provides the most reliable reconstructions but also significantly outperforms previously reported methods. Generative models remain promising but unstable, and iterative solvers offer only marginal improvements, underscoring the persistent ill-posedness of gravity inversion.

		\end{abstract}
		
		
	\end{frontmatter}

	\section{Introduction}		
	
	Exploring the Earth's subsurface geological structures has long posed significant challenges for geoscientists and exploration specialists. Traditional techniques, such as excavation and seismic imaging, are not only costly and time-intensive but also raise concerns about environmental risks \cite{1}. However, recent advancements in geophysical technology have led to the development of more efficient, non-invasive methods, revolutionizing our ability to study the Earth's deep structures. One such method in geophysical exploration is a common technique that uses observed field data to reconstruct subsurface features. Among inversion approaches, gravity inversion is a method used for identifying subsurface anomalies. Our goal is to employ this technique to estimate the density properties of buried objects \cite{2}. By measuring the Earth's gravitational field at specific surface locations, gravity inversion helps identify subsurface density variations, as different masses in the Earth's interior produce gravitational anomalies based on their density contrasts with surrounding materials \cite{3}.
	
	From a mathematical perspective, the inverse problem is the problem of reversing the gravitational forward model, typically represented by the equation $g=A(\rho) $. In this formulation, $g$ corresponds to the measured gravitational data, while $\rho$ denotes the subsurface density distribution \cite{4,5,p}. However, gravity inversion in geophysics presents challenges, primarily due to its ill-posed nature. Since the number of unknown model parameters exceeds the available observed data, the solution becomes non-unique, complicating the inversion process \cite{6,7}. To resolve the ill-posed nature of the problem, one traditional approach is to apply regularization theory \cite{8}. This is achieved by minimizing an objective function that includes a data fitting term, which minimizes the difference between observed and modeled data, and a model constraint term that stabilizes convergence. Several studies have adopted alternative regularization strategies to solve gravity inversion. For instance, total variation regularization using randomized generalized singular value decomposition\cite{9}, wavelet-based multiscale regularization\cite{10}, and a more novel arctangent-based regularization technique\cite{11}. Once the regularization strategy is defined, the next crucial step is choosing an optimization method to minimize the objective function. Deterministic gradient-based methods are widely used in gravity inversion due to their efficiency and well-established mathematical framework\cite{12}. Techniques like steepest descent, conjugate gradient, and Gauss-Newton methods iteratively update the model by following the gradient of the objective function, often enhanced by regularization terms to ensure stability and physical plausibility \cite{13,14,15,16}.  Despite their efficiency, these methods can suffer from sensitivity to the initial model and the risk of getting trapped in local minima. To address these limitations, stochastic and global optimization approaches, such as genetic algorithms \cite{17}, simulated annealing \cite{18}, and particle swarm optimization \cite{19}, have been proposed. These methods explore the solution space more broadly and reduce dependence on initial models, offering improved chances of finding the global minimum at the cost of higher computational effort. To reduce the computational cost, \cite{12} introduces a generalized subspace inversion algorithm that enhances efficiency by working in a lower-dimensional subspace rather than the full model space. Additionally, \cite{20} introduces a wavelet-based multiscale inversion method for gravity data, which addresses efficiency challenges and memory usage by compressing the sensitivity matrix and decomposing the inversion into subproblems of different scales.

	Traditional mathematical inversion techniques often yield limited precision in subsurface reconstructions. This has prompted a shift toward machine learning (ML) approaches\cite{21}, which leverage deep architectures such as CNNs \cite{22}, Physics-Informed Neural Networks (PINNs) \cite{23}, Autoencoders \cite{24}, and GANs \cite{25}.These architectures are particularly effective at handling nonlinearity, high-dimensional data, and complex pattern recognition. In the context of gravity inversion, ML-based methods, particularly CNNs, have shown potential in improving the accuracy and robustness of subsurface modeling due to advantages such as local connections, weight sharing, enhanced generalization capabilities, and transfer learning abilities \cite{26,27}. Similarly,\cite{28} proposes a DL-based approach using an encoder-decoder architecture, where a gravity field encoder extracts 2D gravity features, a transformation module adjusts the feature space, and a density structure decoder reconstructs the 3D density model. Furthermore, U-Nets, with their encoder-decoder design and multi-scale feature extraction, are widely used in gravity inversion. Recent works adapt this architecture to address key geophysical challenges. \cite{29} establishes the baseline use of a standard U-Net architecture for 3D gravity inversion by employing a U-Net-based CNN, demonstrating improved accuracy over traditional least-squares and fully convolutional network (FCN) approaches. To further address challenges such as overfitting and limited generalizability, \cite{30} enhances this framework by introducing an improved U-Net architecture in conjunction with advanced data augmentation techniques (Mixup, flip, noise addition). In contrast, \cite{2} adopts an alternative strategy: rather than refining the U-Net architecture itself, it presents DecNet, a decomposition network that transforms a 2D-to-3D gravity inversion problem into simpler 2D-to-2D mapping process by predicting boundary position, vertical center, thickness, and density distribution , rather than the density of each grid point. \cite{31} expands the scope of U-Net applications by introducing a multitask deep learning framework designed for the simultaneous denoising and inversion of 3D gravity data using a Cross-Dimensional UNet (CDUNet) network. In the following of these efforts,  \cite{7} proposed a stochastic inversion framework using a conditional variational autoencoder (CVAE) to sample the posterior distribution of density models, enabling the integration of prior geological knowledge directly into the inversion process. In a separate development, \cite{6} applies the EfficientNetV2 network for 3D gravity inversion, using composite scaling and Fused-MBConv modules to optimize efficiency, accuracy, and parameter use, achieving superior resolution and stability over traditional methods and other deep learning models like CNNs and ResU-Net++.
	
In addition to purely data-driven approaches, some researchers have explored hybrid methods that integrate physics-based modeling with machine learning techniques to address the limitations of limited prior information and the lack of physical interpretability in data-driven models\cite{32,33}. For instance, \cite{34} reviews Physics-Informed Machine Learning (PIML) approaches for geophysical inversion. It categorizes four PIML strategies (Skeletal, Parallel, Iterative-Sequential, and PINNs), discusses: a unified objective function, overfitting mitigation, transfer learning, which are applied in seismic, gravity, and resistivity inversion. To demonstrate the practical application of these hybrid frameworks, \cite{35} employs a U-Net, incorporating a data-fitting term and synthetic data from random walk simulations to enhance Antarctic crustal density resolution. In an effort to mitigate data scarcity, researchers introduced data-efficient methods such as the Supervised Descent Method (SDM), initially developed by Xiong and Torre \cite{36} for optimization in computer vision. SDM was later adapted for 1D transient electromagnetic inversion \cite{37} and 2D magnetotelluric inversion \cite{38}, demonstrating its effectiveness in electromagnetic applications. advancing this trajectory, \cite{39} introduces an SDM framework for 3D gravity inversion in which the offline phase uses a prior-informed training set to iteratively learn descent directions, and the online phase employs focused regularization and density range restrictions to yield sharp, realistic density models. inspired by these efforts, \cite{40} proposes a Self-Supervised 3D Gravity Inversion (SSGI) framework that eliminates dependency on labeled data by integrating the law of universal gravitation into a closed-loop inversion-forward modeling workflow. Furthermore, \cite{41} advances hybrid workflows with an Enhanced Dual U-Net (EdU-Net) framework, combining two-stage training (pretraining on synthetic models, fine-tuning on target data) with forward-fitting constraints in the loss function.

In Section \ref{Function_setting}, we derive the governing equations for the gravity inversion, which begins with the gravitational potential and Gauss’s law. We then present both the integral and differential formulations of the forward model. These relations form the mathematical foundation for the inverse problem addressed in the subsequent Sections. In \ref{direct_cnn}, we introduce a purely data-driven inversion approach using a CNN, which is trained to map gravity measurements to subsurface densities.  Section \ref{generative_approaches} and \ref{iterative_methods} explore hybrid approaches to gravity inversion that combine data-driven methods with physics of the problem. Section \ref{generative_approaches} focuses on generative models that optimize in the latent spaces of pretrained models, including a baseline generator, a Variational Autoencoder (VAE), and a GAN, to produce density fields that are consistent with observed gravity data. Section \ref{iterative_methods} investigates whether classical and modern iterative optimization methods such as Gradient Descent (GD), Generalized Minimum Residual (GMRES), Loose Generalized Minimum Residual (LGMRES), and Improved Conjugate Gradient (ICG) from \cite{14} can refine CNN-based initial density estimates. 

\section{Governing equations and main results}\label{Function_setting}

	The relation between the gravitational field $\boldsymbol{g}$ and the gravitational potential $\Phi$ is one of the fundamental concepts in gravity modeling. The gravitational field is defined as the negative gradient of the gravitational potential:
	\begin{equation}\label{gravitational field}
		\boldsymbol{g} = -\nabla\Phi
	\end{equation}
	This equation indicates that the direction of the gravitational field always points toward the steepest descent of the potential. From a physical perspective, this means that gravity always pulls objects toward regions of lower potential.
	
	Gauss's law for gravity is a fundamental principle of physics that describes the relationship between the gravitational field and the mass distribution generating it. The law states that the gravitational flux through a closed surface is proportional to the total mass enclosed within that surface. Mathematically, it is expressed as:
	\begin{equation}\label{Gauss's law}
		\oint_{\partial\Omega} \boldsymbol{g}\cdot d\boldsymbol{A} = -{S}_n \gamma\displaystyle\int_{\Omega} \rho \, d\boldsymbol{x}
	\end{equation}
	In this expression, $\boldsymbol{g}(\boldsymbol{x})$ is the gravitational field at point $\boldsymbol{x}$, $d\boldsymbol{A}$ is an outward-pointing surface element on the closed surface $\partial \Omega$, $S_n$ denotes the surface area of the unit sphere in $n$ dimensions, $\rho(\boldsymbol{x})$ is the mass density at point $\boldsymbol{x}$, $\gamma$ is the universal gravitational constant, and $\Omega$ is the volume enclosed by the surface $\partial \Omega$. This law provides the foundation for linking gravity measurements to subsurface mass distribution, forming the core of gravitational inverse problems.
	
	Assuming the gravitational field $\boldsymbol{g}$ is continuously differentiable and applying the divergence theorem to Gauss's law (\ref{Gauss's law}) yields:
	\begin{equation}\label{Gauss-div}
		\int_{\Omega} \text{div}(\boldsymbol{g})\, d\boldsymbol{x}  = -{S}_n \gamma\int_{\Omega} \rho \, d\boldsymbol{x} 
	\end{equation}
	Since this relation holds for any arbitrary volume $\Omega$, it follows that:
	\begin{equation}\label{gravitational eq}
		\text{div}(\boldsymbol{g}) = -{S}_n \gamma \rho
	\end{equation}
	Substituting $ \boldsymbol{g} = -\nabla\Phi$ into the above differential equation leads to the well-known Poisson’s equation:
	\begin{equation}\label{Poisson}
		\nabla^2\Phi = {S}_n \gamma\rho
	\end{equation}
	
	To solve this equation, we utilize Green's functions corresponding to the Laplace equation ($\nabla^2\Phi = 0$). Since the gravitational potential $\Phi(\boldsymbol{x})$ depends only on the distance between points, we aim to find radial solutions. Thus, the Green’s functions in two and three dimensions are given by:
	\begin{equation}\label{Green's functions}
		\mathbb{G}(\boldsymbol{x}, \boldsymbol{x}^{\prime}) = 
		\begin{cases}
			\phantom{-}\dfrac{1}{2\pi} \ln \left\lvert \boldsymbol{x} - \boldsymbol{x}^{\prime} \right\rvert + C_1\,\,\, , &\qquad \boldsymbol{n} = 2 \\\\
			-\dfrac{1}{4\pi | \boldsymbol{x} - \boldsymbol{x}^{\prime} |} + C_2 \,\,\, ,&\qquad\boldsymbol{n} = 3
		\end{cases}
	\end{equation}
	
	Assuming that the mass density $\rho$ has compact support (i.e., $\rho = 0$ outside a bounded region $\Omega$), the gravitational potential at a point $\boldsymbol{x}$ can be expressed using relation (\ref{Green's functions}) as:
	\begin{equation}\label{potential_fields}
		\Phi(\boldsymbol{x}) = 
		\begin{cases}
			\phantom{-} \gamma \displaystyle\int_{\Omega} \rho(\boldsymbol{x}^{\prime}) \ln \left| \boldsymbol{x}-\boldsymbol{x}^{\prime} \right| d\boldsymbol{x}^{\prime} + C_1^{\prime}, &\qquad \boldsymbol{n} = 2 \\\\
			- \gamma \displaystyle\int_{\Omega} \frac{\rho(\boldsymbol{x}^{\prime})}{\left| \boldsymbol{x}-\boldsymbol{x}^{\prime} \right|} d\boldsymbol{x}^{\prime} +C_2^{\prime}, &\qquad \boldsymbol{n} = 3
		\end{cases}
	\end{equation}
	
	As a result, the gravitational field $\boldsymbol{g}$ can be computed by combining equations \ref{potential_fields} and \ref{gravitational field}:
	\begin{equation}\label{gravity}
		\boldsymbol{g}(\boldsymbol{x}) = 
		\begin{cases}
			-\gamma \displaystyle\int_{\Omega} \rho(\boldsymbol{x}^{\prime}) 
			\frac{\boldsymbol{x}-\boldsymbol{x}^{\prime}}{\lvert \boldsymbol{x}-\boldsymbol{x}^{\prime} \rvert^2} \, d\boldsymbol{x}^{\prime}, &\qquad \boldsymbol{n} = 2 \\\\
			-\gamma \displaystyle\int_{\Omega} \rho(\boldsymbol{x}^{\prime}) 
			\frac{\boldsymbol{x}-\boldsymbol{x}^{\prime}}{\lvert \boldsymbol{x}-\boldsymbol{x}^{\prime} \rvert^3} \, d\boldsymbol{x}^{\prime}, &\qquad \boldsymbol{n} = 3
		\end{cases}
	\end{equation}

	In the forward problem, when the mass density $\rho(\boldsymbol{x})$ is known, the goal is to compute the gravitational field $\boldsymbol{g}$ using equation \ref{gravity}. In contrast, the gravity inversion problem aims to recover the unknown mass density $\rho(\boldsymbol{x})$ within a region $\Omega$ based on measured gravitational data.

	This paper focuses on the two-dimensional formulation of the gravity inversion problem. For numerical simulation, the domain $\Omega$ is discretized into $N = n \times n$ individual cells, and it is assumed that gravitational measurements are taken at $n$ points on the surface. Since the relationship between density and the gravitational field is linear (as expressed in equation \ref{gravity}), this relationship can be formulated as a linear system of equations:
	$$(A_{2n \times n^2}) \rho_{n^2 \times 1} = {g}_{2n \times 1}$$
	Note that, due to the presence of both horizontal and vertical components of the gravitational field, the total number of gravity measurements in the above system is $2n$.

    Given the data-driven nature of the proposed methods, we employ a synthetic dataset for validation. This dataset comprises artificial density models introduced in our prior work \cite{42}, along with their associated gravitational fields, calculated using equation \ref{gravity}.

	 In the following Sections, we present the modeling details, the inversion framework, and the implementation of machine learning methods for density reconstruction.


	\section{First method (Direct Inversion Using CNNs)}\label{direct_cnn}
	
\begin{figure}
		\centerline{\includegraphics[scale=0.70]{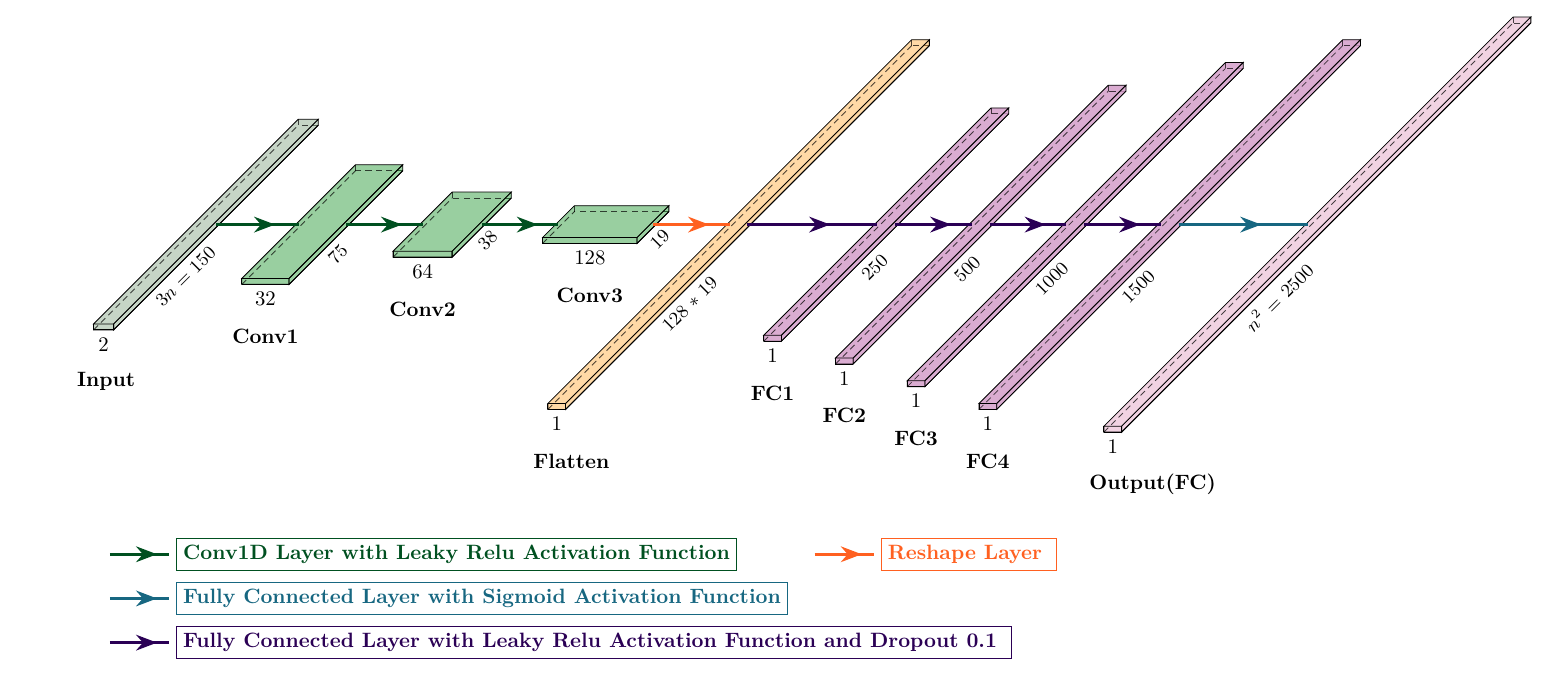}}
		\caption{The proposed CNN architecture is designed for the direct inversion of gravity data. A total of $3n$ gravity measurements (with $n=50$) are collected from the surface to estimate the underlying density. Since each gravity measurement includes both horizontal and vertical components, the input tensor width is set to 2.
		}
		\label{CNN}
	\end{figure}

\begin{figure}
		\centerline{\includegraphics[scale=0.70]{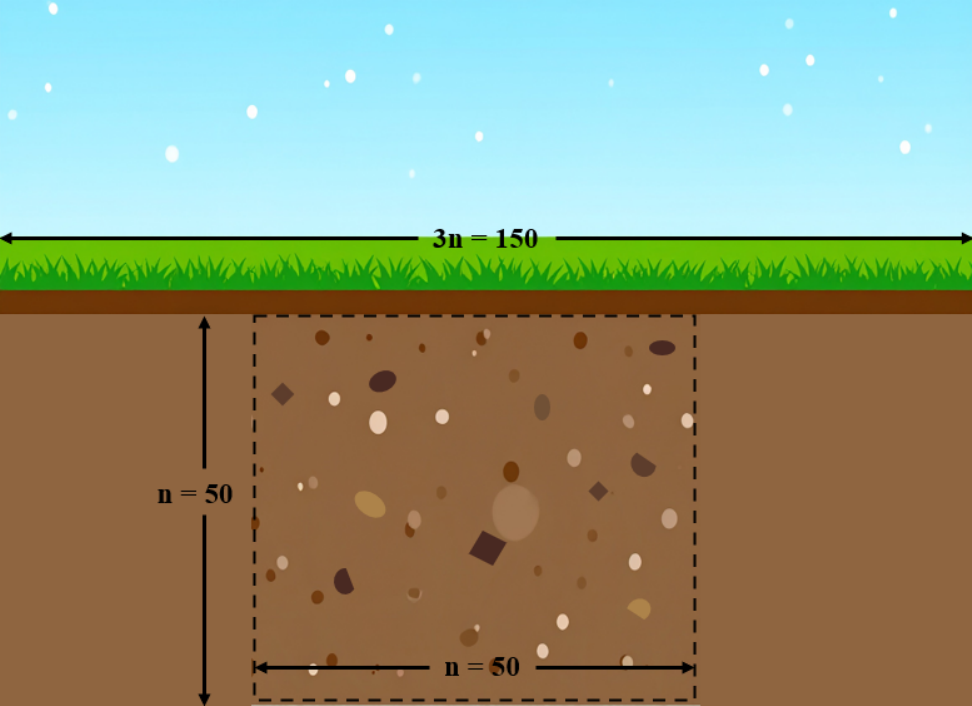}}
		\caption{A schematic representation of the gravity data configuration is shown. A total of $3n = 150$ gravity measurements are collected along the surface to capture the horizontal distribution of the gravity anomaly. The subsurface region (density) is discretized into a grid of size $n \times n = 50 \times 50$, as indicated by the dashed box.
		}
		\label{ground}
	\end{figure}

	As the first of our data-driven approaches, we employ a direct inversion method based on a CNN to estimate the subsurface density distribution from gravity anomaly measurements. To achieve this, we design and train a CNN that takes a 1D surface gravity anomaly vector as input and outputs the corresponding 2D subsurface density distribution. Although equation \ref{gravity} defines a linear relationship between gravity and density in the form $g=A\rho$, we solve the inverse problem nonlinearly by training the network to learn the mapping from gravitational measurements to the underlying density field.
	 The structure of the CNN undergoing training is depicted in Figure \ref{CNN}.
	 As a novel enhancement, we found that increasing the number of gravity observation points from $n $ to $ 3n$ significantly reduces inversion error. The expanded volume of input data to the network produces more accurate and stable density estimates. A schematic illustration of the data acquisition setup is shown in Figure \ref{ground}, where surface gravity measurements are collected above a discretized subsurface domain. 

	\section{Second method (Generative Approaches to Gravity Inversion)}\label{generative_approaches}
	This Section introduces the use of generative models in the context of gravity inversion. In the first step, a generative network is designed to reconstruct real subsurface density fields (using generators such as VAEs or GANs). The generator takes noise following a specific distribution and outputs a density field with dimensions identical to those in the dataset. After training, we treat the generator as a nonlinear function $ G : \mathbb{R}^k \rightarrow \mathbb{R}^{n^2}$ , where $\boldsymbol{z}\in \mathbb{R}^k$ is a latent vector and $G(\boldsymbol{z})$ is the corresponding density model in the physical domain. This representation enables us to reformulate gravity inversion as an optimization problem in the generator's latent space. The objective is to find a latent vector $\boldsymbol{z} \in \mathbb{R}^k$ such that generated density $G (\boldsymbol{z})$ satisfies the forward model equation. That is the relationship  $ AG(\boldsymbol{z})\approx g$ must hold, where $A$ is the known forward operator and $g$ is the observed gravity data. Specifically, we aim to minimize the Euclidean mismatch between observed and predicted gravity data:
	\begin{equation}\label{gravity_approx}
		\boldsymbol{z}^{*} = \arg\min_{z} \left\|AG(\boldsymbol{z})-g \right\|_2^2\cdot
	\end{equation}
	Since the function G is nonlinear, we employ gradient-based optimization methods to update $\boldsymbol{z}$ for solving \ref{gravity_approx}. If G were a linear function, equation \ref{gravity_approx} would reduce to a standard least squares problem. However, in practice, a linear generator cannot reconstruct realistic density fields. Subsequently, we provide a concise explanation of the two generators used in this research.

\begin{figure}
	\centerline{\includegraphics[scale=0.70]{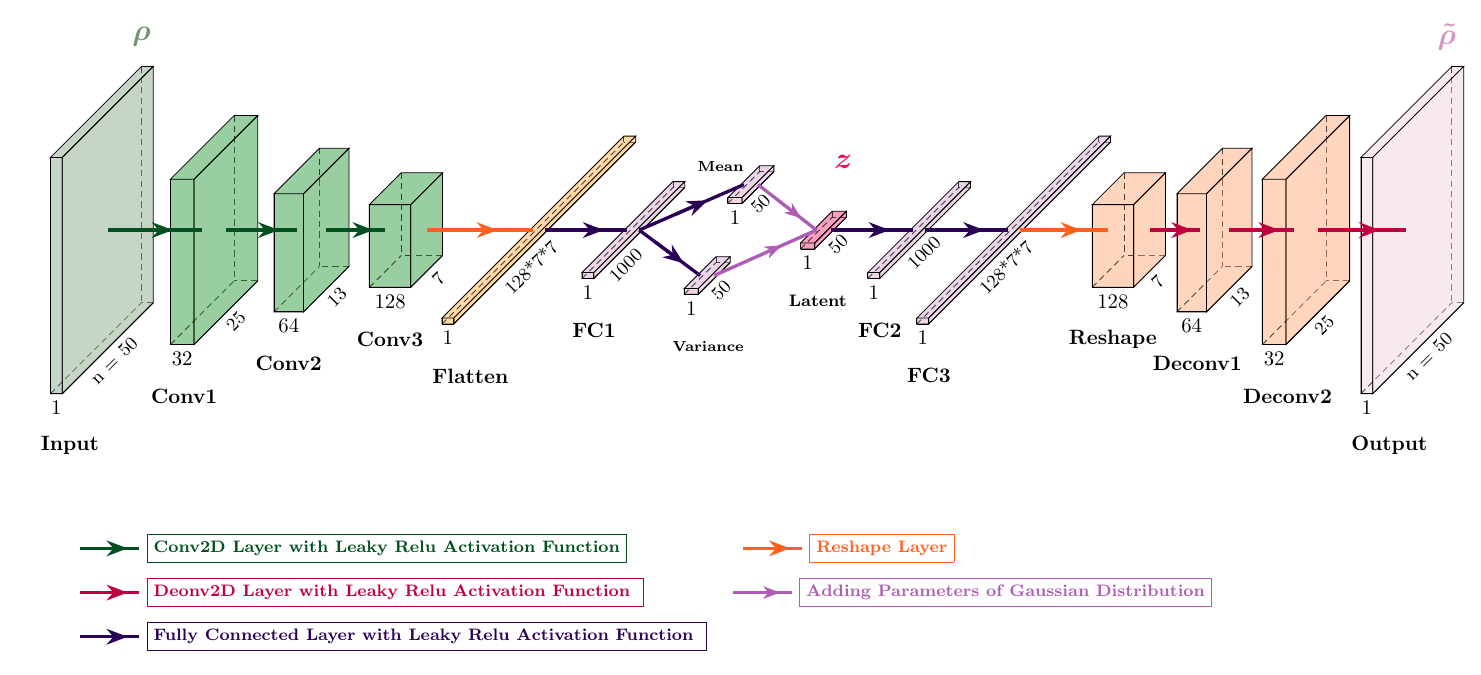}}
	\caption{Architecture of the VAE used for gravity inversion, where the decoder serves as a generator in latent-space optimization}
	\label{VAE_fig}
\end{figure}

\begin{figure}[H]
	\centerline{\includegraphics[scale=0.55]{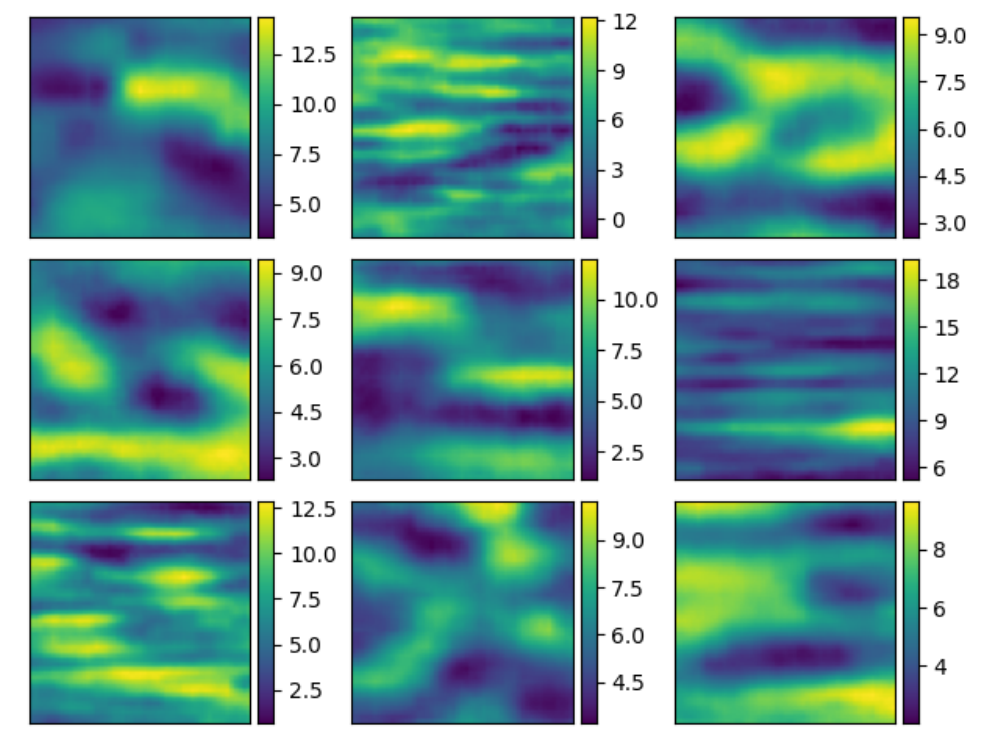}}
	\caption{Random sample generation from the VAE}
	\label{Gen_VAE}
\end{figure}

\paragraph{\textbf{Variational Autoencoders (VAE)}}\label{VAE} 
VAEs are the probabilistic generative models designed to learn latent representations of input data in a lower-dimensional space and reconstruct them from this space. These networks comprise two core components: an encoder, which maps input data to the latent space, and a decoder, which reconstructs the original data from the latent vector. The loss function used during training is defined as follows:

\begin{equation}
	\mathcal{L}_{\text{VAE}} = \mathbb{E}_{q_\phi(\boldsymbol{z}|\boldsymbol{x})} [-\log p_\theta(\boldsymbol{x}|\boldsymbol{z})] + \mathcal{D}_{KL}[ q_\phi(\boldsymbol{z}|\boldsymbol{x}) \ ||\ p(\boldsymbol{z})]
\end{equation}
where $\boldsymbol{x}$ is the input (a true density), $\boldsymbol{z}$ is a latent vector, $q_{\phi}$ is the encoder, and $p_{\theta}$ is the decoder. The first term encourages accurate reconstruction of data, while the second term minimizes the divergence between the learned posterior and normal distribution. Once the training is complete, we discard the encoder and use the decoder as a mapping from latent to physical space. We then optimize over variable $\boldsymbol{z}$ to find a density field that best fits the observed gravity data through the objective function defined in equation \ref{gravity_approx}. The architecture of the VAE used in this approach is illustrated in Figure \ref{VAE_fig}. Additionally, Figure \ref{Gen_VAE} displays randomly sampled outputs from the trained generator, demonstrating its capability to effectively learn and reconstruct the hidden structures present in the real data.

\begin{figure}[H]
	\centerline{\includegraphics[scale=0.70]{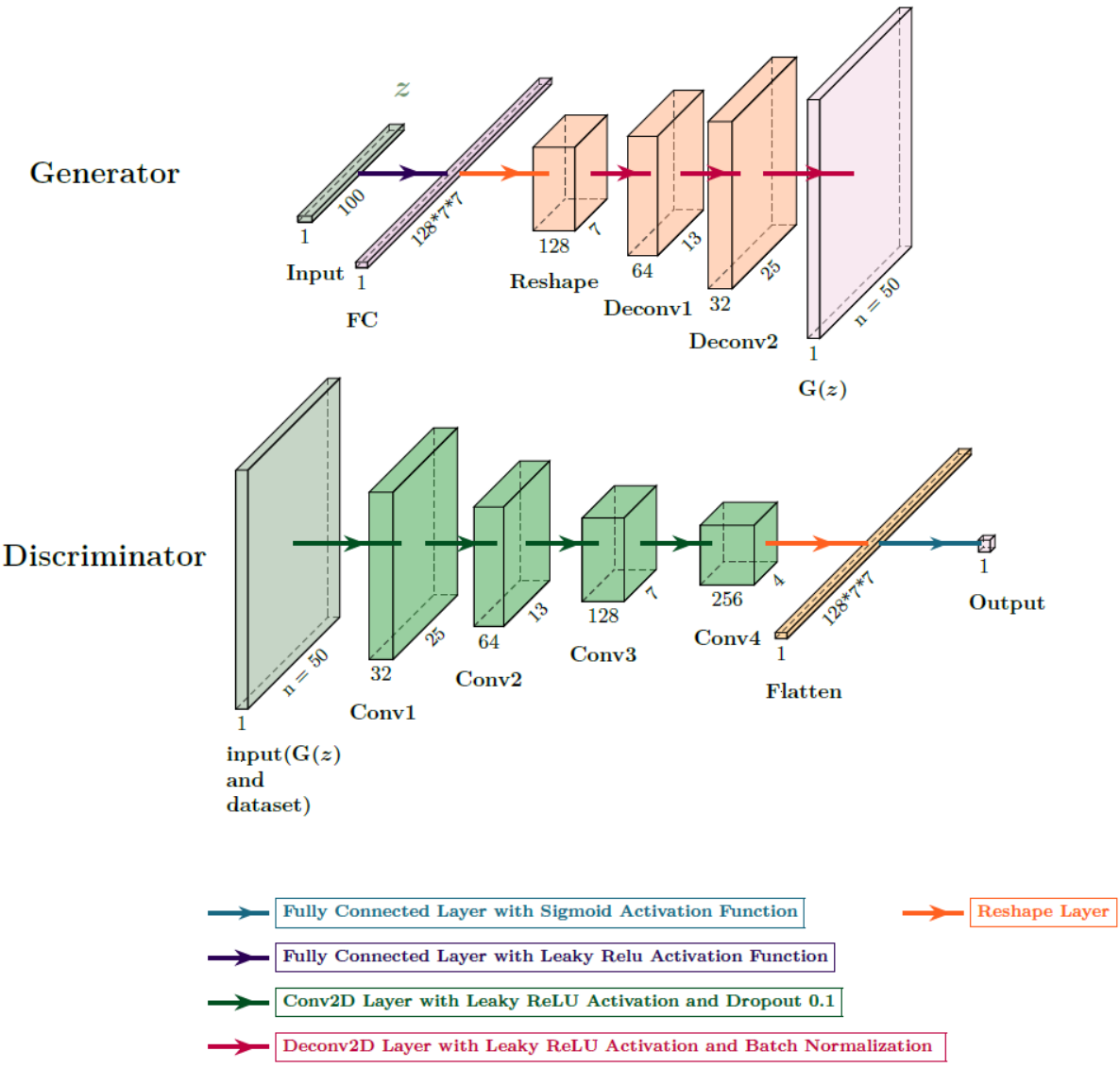}}
	\caption{Architecture of the GAN used for gravity inversion, where the trained generator is later used for latent-space optimization}
	\label{GAN_fig}
\end{figure}

\begin{figure}[H]
	\centerline{\includegraphics[scale=0.45]{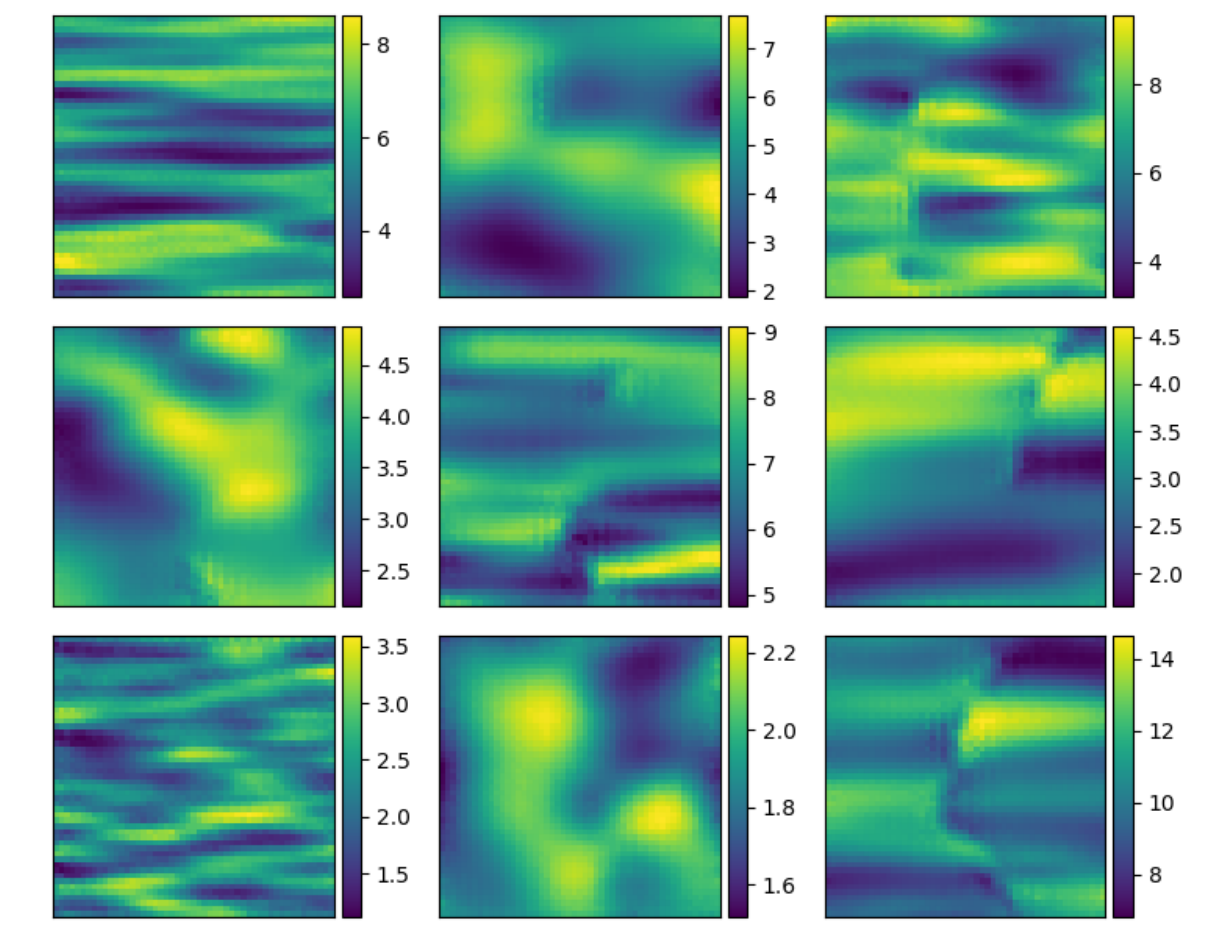}}
	\caption{Random sample generation from the GAN}
	\label{Gen_GAN}
\end{figure}

\paragraph{\textbf{Generative Adversarial Network (GAN)}} \label{GAN}
    GANs consist of two neural networks: a generator G and a discriminator D, trained in an adversarial setting. The generator learns to produce realistic samples from random latent vectors, while the discriminator tries to distinguish between real data and the generator’s output. The standard GAN loss function is defined as a minimax optimization problem:
\begin{equation}
	\min_G \max_D \, \Big( \mathbb{E}_{\boldsymbol{x}}  [\,\log D(\boldsymbol{x})\,] + \mathbb{E}_{\boldsymbol{z}}  [\,\log(1 - D(G(\boldsymbol{z})))\,] \Big)
\end{equation}
    Here, $\boldsymbol{x}$ is a real data sample (true density), and $\boldsymbol{z}$ is a latent vector sampled from a prior distribution (typically Gaussian). The discriminator D is trained to assign high probability to real data and low probability to fake data G($\boldsymbol{z}$), while the generator G is trained to fool the discriminator. After full training, the discriminator network is discarded, and the generator is employed to reconstruct density fields. Similar to the VAE approach, we optimize the latent vector $\boldsymbol{z}$ using objective function \ref{gravity_approx} to yield density fields that reconstruct the observed gravity data. Figure \ref{GAN_fig} illustrates the complete GAN architecture, highlighting the inversion-stage generator structure.
    Additionally, Figure \ref{Gen_GAN} displays randomly sampled outputs from the trained generator, demonstrating its capability to effectively learn and reconstruct the hidden structures present in the real data.
    
\begin{figure}[H]
	\centerline{\includegraphics[scale=0.5]{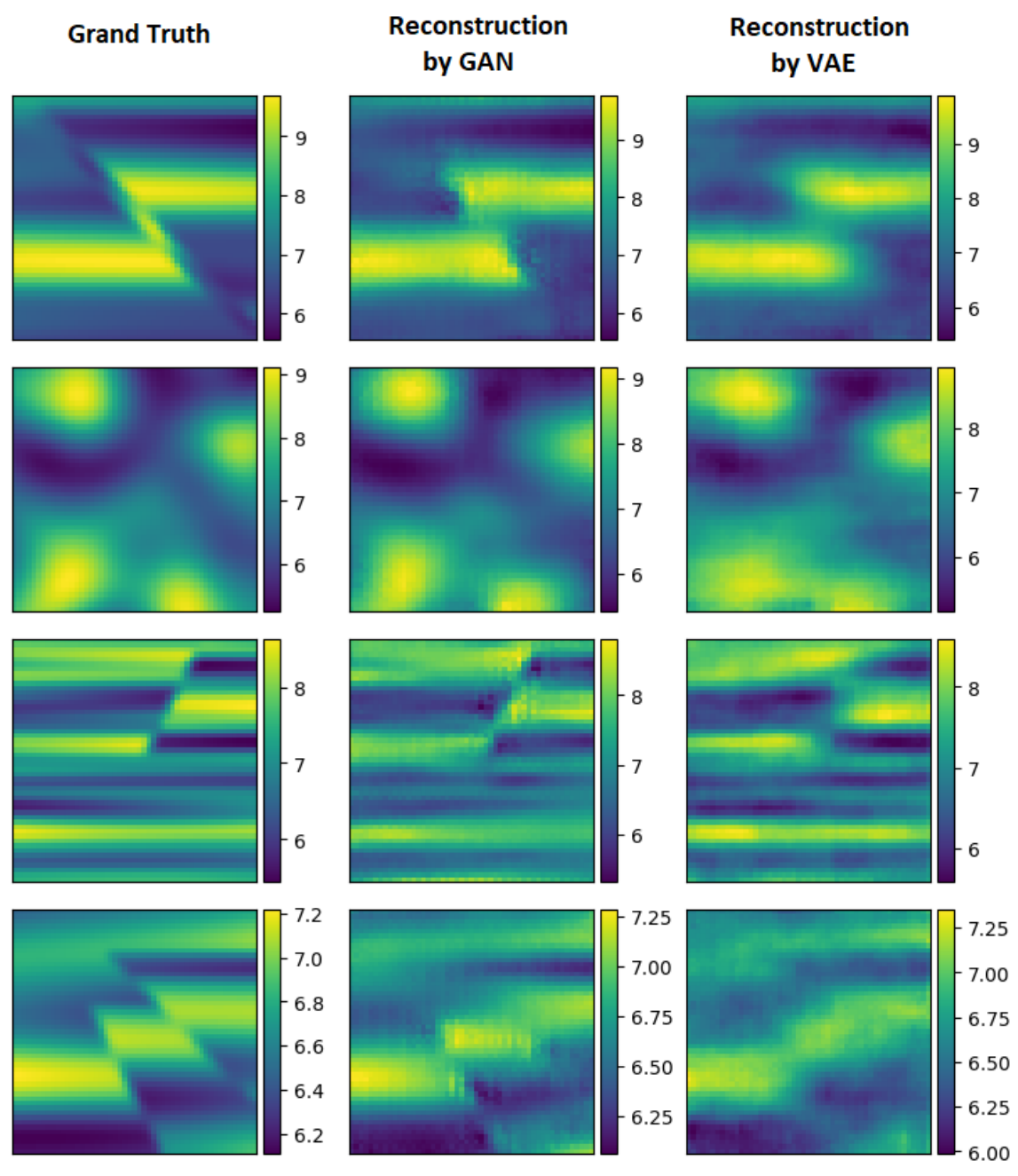}}
	\caption{Reconstruction of the true density using the generators. The latent vector $\boldsymbol{z}$ is obtained by solving the optimization problem $||G(\boldsymbol{z}) - \rho||_2^2$, where $G(\boldsymbol{z})$ serves as an approximation of the true density.}
	\label{Generator_Power}
\end{figure}

\paragraph{\textbf{Key Considerations During Inversion:}}
When performing gravity inversion using trained generative models, the following points are critical:
\begin{enumerate}
	\item 
	These generators can produce density fields with geologically realistic physical properties. To demonstrate this capability, refer to Figure \ref{Generator_Power}. In this experiment, we selected several density fields from the test dataset (denoted as $\rho$). We then performed optimization to find latent vector $\boldsymbol{z}$ that minimizes the $||G(\boldsymbol{z}) - \rho||_2^2$ error. As shown in Figure \ref{Generator_Power}, the generators successfully reconstruct these density fields, indicating effective learning of the true density distribution.
	
	\item 
	This finding raises the prospect of obtaining realistic estimates of $\rho$ through minimization of the $||AG(\boldsymbol{z}) - g||_2^2$ error function with respect to $\boldsymbol{z}$ using observed gravity data $g$, as outlined in the beginning of the Section \ref{generative_approaches}. However, Figure \ref{Generator_Result_Gravity} reveals that optimizing this error function with different initial values for $\boldsymbol{z}$ converges to entirely distinct density models, all of which exhibit valid physical structures, and significantly reduce the error metric. This behavior demonstrates the ill-posedness of the equation  $A\rho =g$, even within the space of solutions with meaningful physical characteristics.
	
	\item 
	 Given the issue raised in the second paragraph and considering the desirable property outlined in the first paragraph, a promising approach was to first find a latent vector $\boldsymbol{z}_0$  such that $G(\boldsymbol{z}_0)$ closely approximates the density estimate $\rho$ obtained via CNN (First method). This $\boldsymbol{z}_0$ then serves as the initial value for optimizing the $||AG(\boldsymbol{z}) - g||_2^2$ error function in latent space. However, as observed in Figure \ref{Generator_Initial}, the final solution shows no striking deviation from the initial estimate, reducing the error function only through minor adjustments.
\end{enumerate}

	\section{Third method (CNN-Based Initialization for Iterative Optimization)}\label{iterative_methods}

	Iterative methods are techniques that aim to converge toward an optimal solution by starting from an initial guess and successively updating it. In the context of gravity inversion, the objective of these methods is to improve the initial estimate such that the resulting model is consistent with the governing physical equations (i.e., the forward model) and matches the observed data. In this case, the forward model is represented by a linear system.
	
	As described in Section \ref{direct_cnn}, a CNN has been trained to directly predict the subsurface density field from surface gravity data. In this Section, we investigate whether combining this learned initial estimate with classical and modern iterative solvers can enhance inversion accuracy.
	Specifically, we use the CNN output as a starting point and apply a set of iterative methods including GD, GMRES, LGMRES, as well as the ICG method proposed in \cite{14} to solve the linear system $A\rho = g$.

	 \begin{remark}
	 	 It is worth noting that Equation $A\rho = g$ can be formally solved using a regularized least squares approach, which does not require an initial estimate. However, such a solution typically deviates significantly from the true subsurface structure and is therefore not presented in this work. A detailed discussion of this limitation can be found in \cite{8, 9}.
	 \end{remark}
	  
	 \paragraph{\textbf{GD solver}} This method is an iterative optimization algorithm, which is effective when a good initial estimate is available. In this approach, the linear system $A\rho = g$ is solved by iteratively updating the estimates in the direction opposite to the gradient of the error function, aiming to minimize the least squares objective. The update rule at each iteration is defined as follows:
	 \begin{equation}
	 	\rho^{(k+1)} = \rho^{(k)} - \eta \nabla \mathcal{L}(\rho^{(k)})
	 \end{equation}
	 where $\eta$ is the learning rate and $\mathcal{L}$ denotes the loss function $\mathcal{L}(\rho) = \lVert A\rho - g \rVert_2^2$. 
	 Moreover,
	 $\rho^{(0)}$
	 is considered as an initial guess for the density.
	  
	\begin{remark}
		In our setup, the forward model matrix $A$ is not square, which makes direct application of many solvers challenging. To address this, we convert the system $A\rho = g$ into a square form in such a way that:
		\begin{equation}\label{normal_equation}
			A^T A \rho = A^T g \quad \Longrightarrow \quad B\rho = \hat{g}
		\end{equation}
		where in the above equation, $B_{n^2 \times n^2} = A^T A$ and $\hat{g}_{n^2 \times 1} = A^T g$. This relationship is referred to as the normal equation. Such reformulation ensures compatibility with all solvers considered, including GMRES, LGMRES, and ICG, which allows us to apply them consistently across experiments.
	\end{remark}

	\paragraph{\textbf{GMRES solver}} This is an iterative method designed for solving high-dimensional linear systems. At each iteration, it minimizes the residual magnitude.
	GMRES employs the Arnoldi process to construct an orthonormal basis ${Q=[q_1, q_2, \dots, q_k]}$ for the Krylov subspace $ {\mathcal{K}_k(A, r^{(0)}) = \mathrm{Span}\{r_0, Br_0, \dots, B^{k-1}r_0\}} $, satisfying the relation $BQ = QH$, where $r^{(0)} = {\hat{g}} - B\rho^{(0)}$ is the first residual computed from the initial guess and $H$ is a Hessenberg matrix. Within the Krylov subspace, the goal is to find a vector that minimizes the residual when added to the initial guess. Formally, we seek $y$ such that with the definition of $\rho^{(1)} = \rho^{(0)} + Qy$, minimizes the Euclidean residual norm $\|r^{(1)}\|_2$. This is equivalent to solving the following optimization problem:
	\begin{equation*}
		\begin{split}
			||r^{(1)}||_2^2 &= \| \hat{g} - B\rho^{(1)} \|_2^2  
			= \|r^{(0)} - BQy\|_2^2 = \|Bq_1 - QHy\|_2^2 \\
			&= \|Q(Be_1 - Hy)\|_2^2 = \|Be_1 - Hy\|_2^2
		\end{split}
	\end{equation*}
	Solving this equation via least squares yields vector $y$, and then $\rho^{(1)}$ is computed. If $\rho^{(1)}$  does not provide an adequate approximation of the solution, the algorithm can be restarted using the updated residual $r^{(1)}$ and refined guess $\rho^{(1)}$, continuing iteratively.

	\paragraph{\textbf{LGMRES solver}} This algorithm is an extended version of classic GMRES that accelerates convergence by augmenting the Krylov subspace with solution approximation vectors from prior iterations. This technique has demonstrated superior capabilities for ill-conditioned systems.
    
    \paragraph{\textbf{ICG solver}}
This method is an extension of the classical conjugate gradient (CG) algorithm. In the CG method, conjugate directions $v_k$ are constructed with respect to a given matrix $B$, and the iterative sequence proceeds in the space spanned by these directions according to the update rule ${\rho^{k+1} = \rho^{k} + \eta v_k}$. The enhanced version, ICG, implements key optimizations:  vectorization of diagonal weighting matrices to reduce memory overhead, reformulating gradient computations to avoid explicit matrix construction, and a simplified update scheme that minimizes memory usage by avoiding storage of redundant intermediate vectors.


	\section{Results}\label{Results}	
	
The dataset was divided into training and test sets, containing 27,000 and 1,000 samples, respectively, and the numerical and visual results of the proposed methods were obtained accordingly.
Table \ref{table1} presents the accuracy of the proposed approaches in estimating the density function in this paper. In this table, the symbol $\Omega$ denotes the region defined by the specified coordinate range.
Tables \ref{table2} and \ref{table3} are dedicated to evaluating the performance of the auxiliary networks (i.e., the generative models VAE and GAN).
It is worth noting that in all presented tables, the reported error values represent the average errors computed over the corresponding dataset (either training or test).

This paper aims to identify the optimal and most accurate network architecture for training through a series of experiments.
In training all networks, a batch size of 25 and the Adam optimizer were used.  
Each network also employed its own number of epochs and learning rate decay schedule.  
The exact values of the hyperparameters are available in the provided code.

\begin{figure}[H]
		\centerline{\includegraphics[scale=0.6]{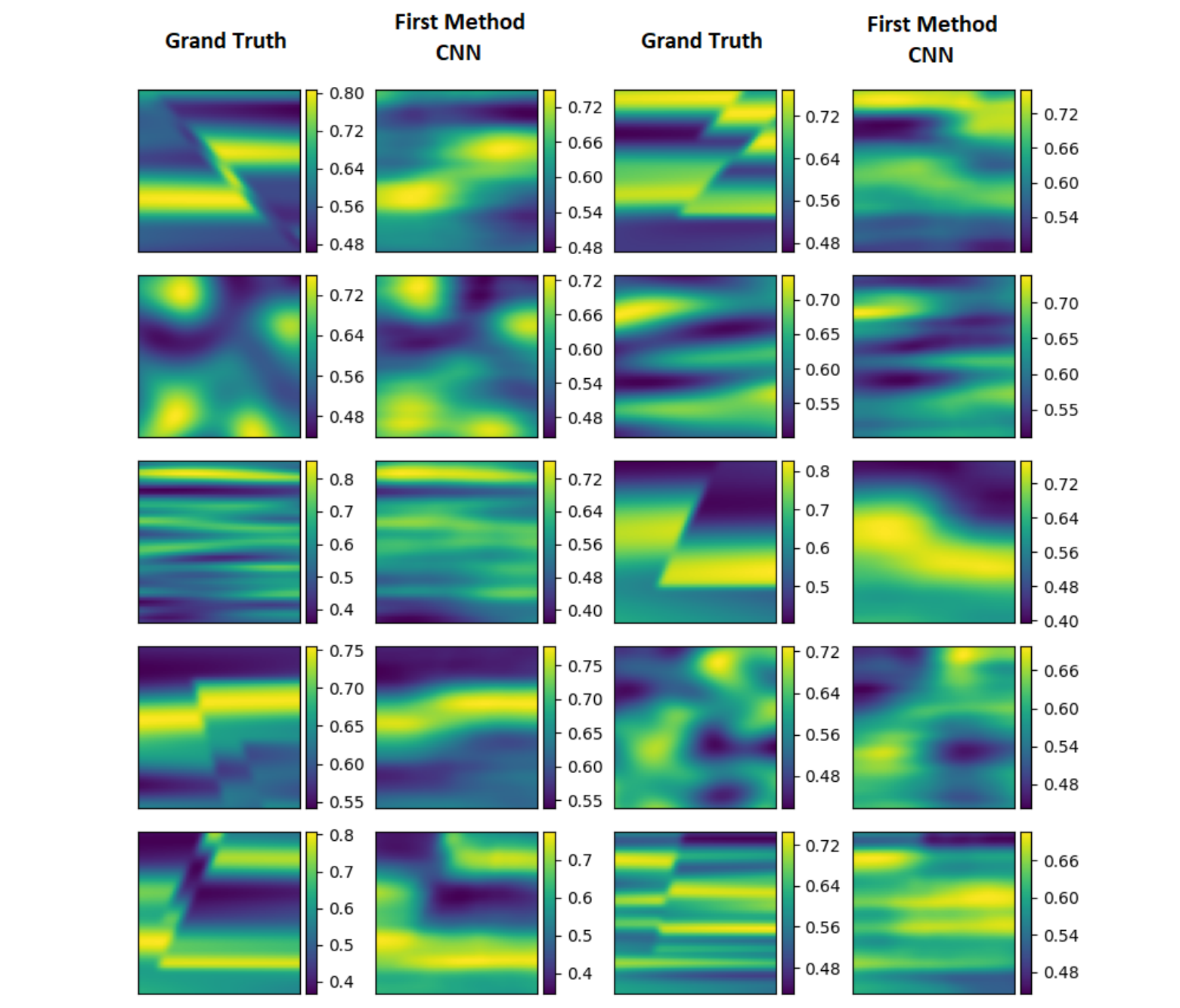}}
		\caption{
			Comparison between the results of the First method and the true density 
		}
		\label{CNN_Gravity}
	\end{figure}

\begin{figure}[H]
		\centerline{\includegraphics[scale=0.40]{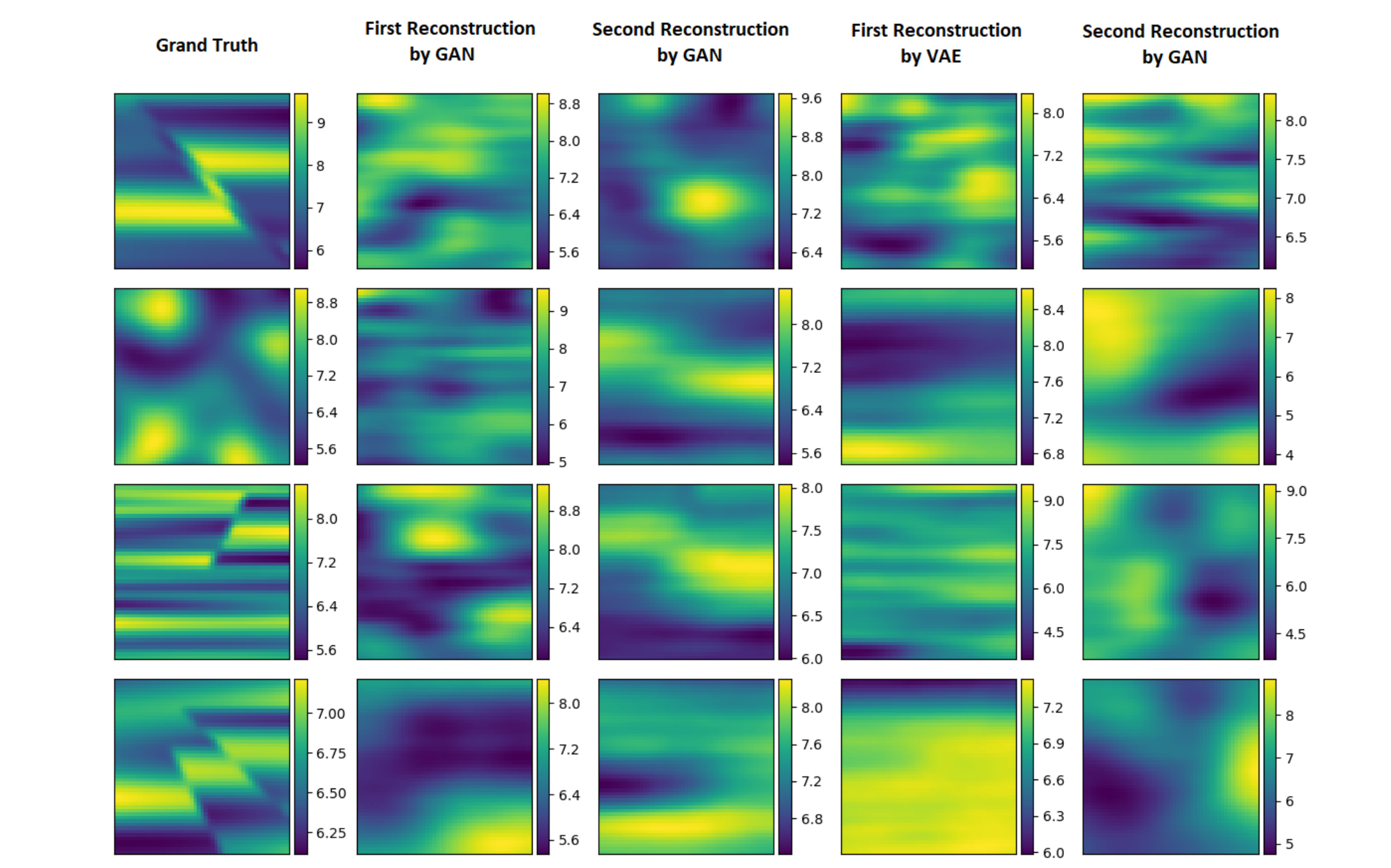}}
		\caption{
			Comparison between the results of the Second method 
			(generator-based approach)
			and the true density. Two different initial guesses (noise vectors) for the latent space lead to two distinct density estimations.
		}
		\label{Generator_Result_Gravity}
	\end{figure}

\begin{figure}[H]
		\centerline{\includegraphics[scale=0.45]{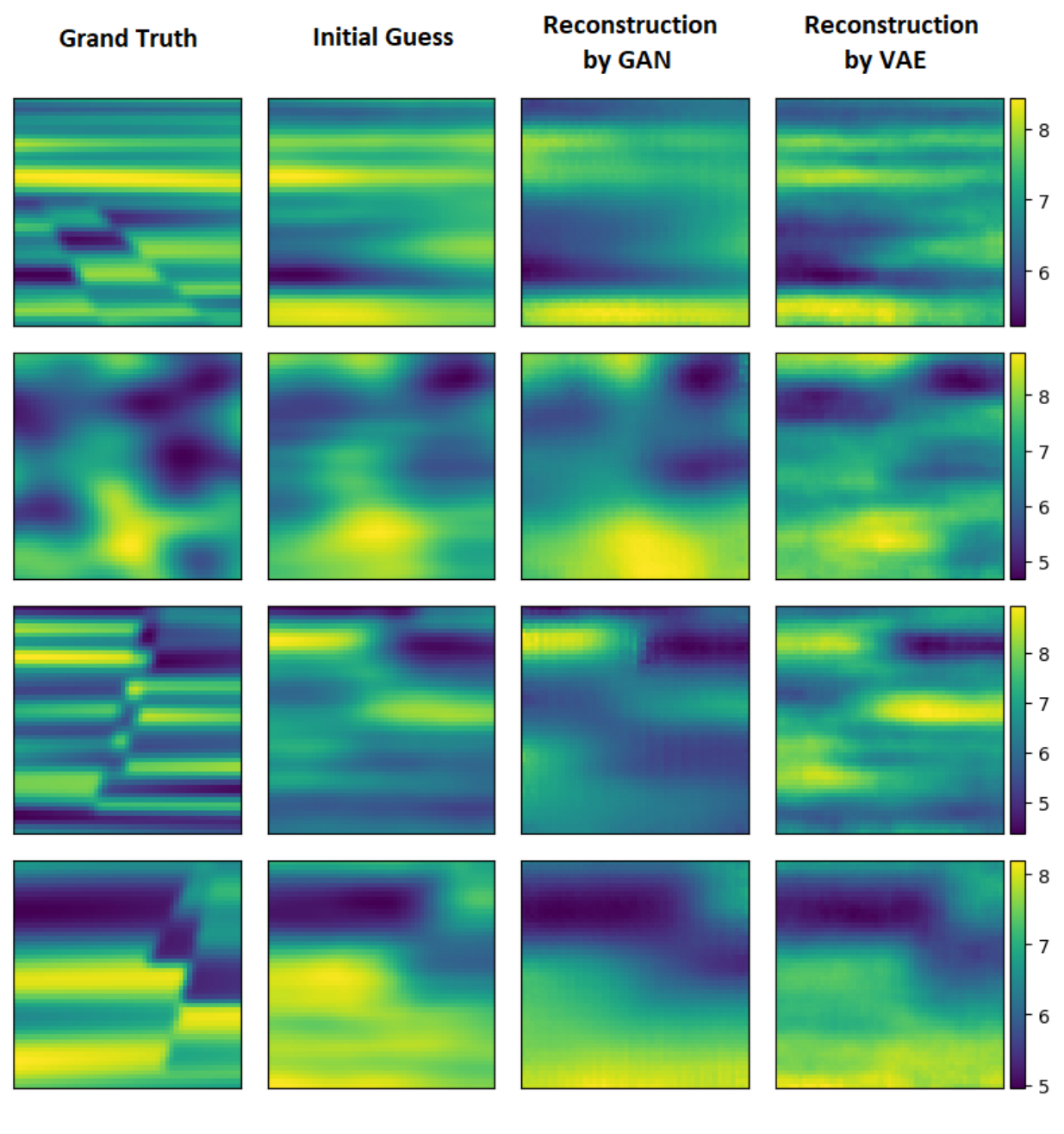}}
		\caption	{
			Reconstruction of the true density using generators initialized with the estimate obtained from the First method.
		}
		\label{Generator_Initial}
	\end{figure}

\begin{figure}[H]
		\centerline{\includegraphics[scale=0.40]{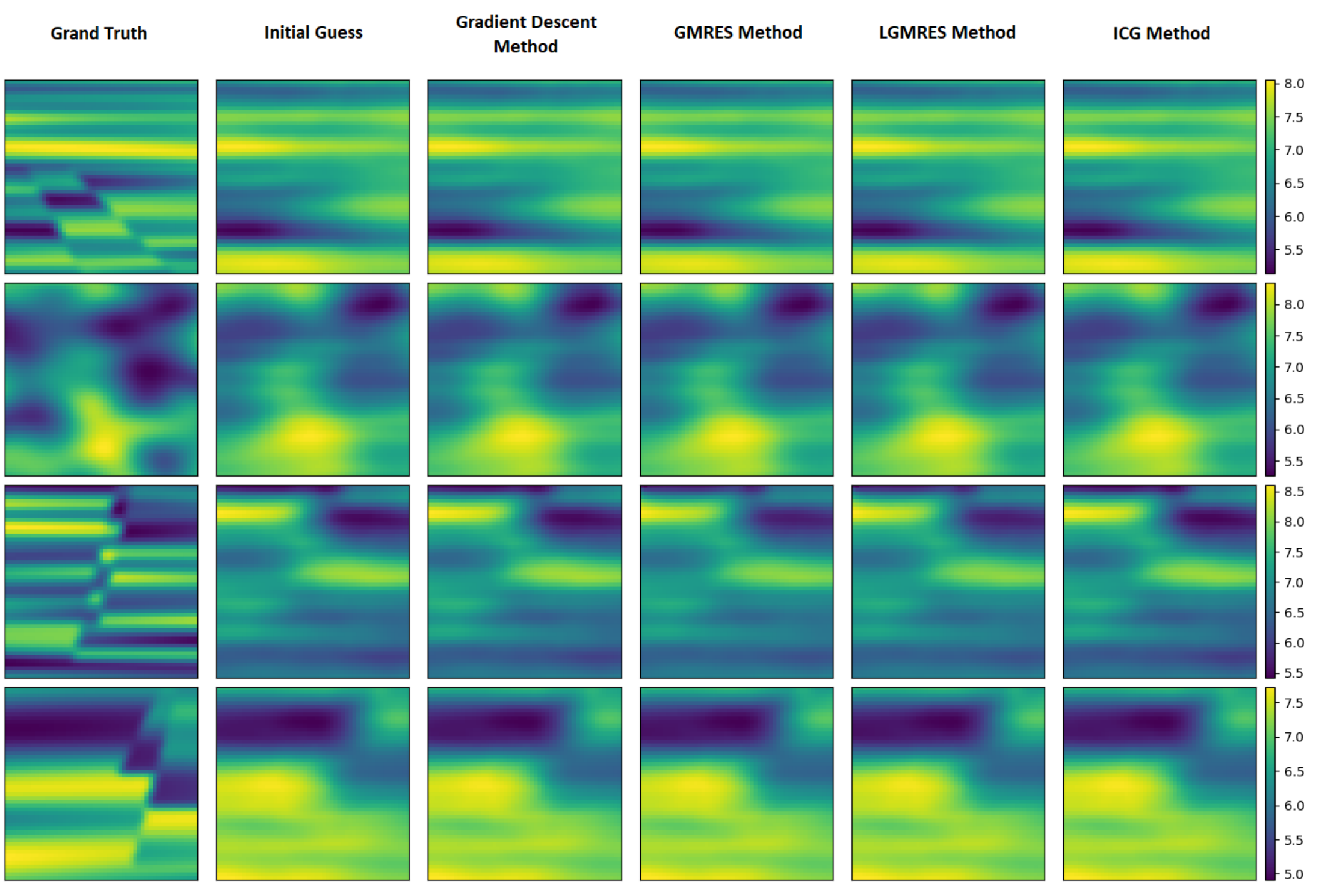}}
		\caption{
			Comparison between the results of the Third method  
			(iterative methods)  
			and the true density.
		}
		\label{Iterative_Gravity}
	\end{figure}

A visual comparison between the different methods is illustrated in Figures  
\ref{CNN_Gravity},  
\ref{Generator_Result_Gravity},  
\ref{Generator_Initial},  
and \ref{Iterative_Gravity},  
which show the results obtained on the test dataset.  
Figures \ref{CNN_Gravity} and \ref{Iterative_Gravity} provide a comparison between several samples of grand truth  density fields and the estimates obtained by the First method (CNN) and the Third method (iterative approach), respectively.  
Moreover, Figures \ref{Generator_Result_Gravity} and \ref{Generator_Initial} compare the ground truth density fields with the results produced by the second method (generator-based approach).  
It should be noted that in Figure \ref{Generator_Initial}, the initial estimate used was generated by the first method.	

In all iterative methods, the main system of equations,
$A\rho = g$,
was reformulated as the normalized equation \ref{normal_equation}, except for GD. In addition, the output of the first method (CNN) was used as the initial guess for each solver.
 The implementation details of each method are described below.

\begin{itemize}
	
\item 
\textbf{GD:} Solves the system using a learning rate of $10^{-5}$, a stopping tolerance of $10^{-6}$, and a maximum of 100 iterations.

\item 
\textbf{GMRES:} Implemented using SciPy's gmres solver with a relative tolerance of $10^{-6}$, no absolute tolerance, and up to 5000 iterations.

\item 
\textbf{LGMRES:} Based on SciPy’s lgmres solver,  using an inner Krylov subspace size of 30, 50 outer vectors, a relative tolerance of $10^ {-6}$, and a maximum of 5000 iterations.
\item 
\textbf{ICG:} A regularized iterative conjugate gradient method that applies data and model weighting, dynamically adjusts the regularization parameter based on initial misfit values, and reduces it with a decay factor $ q=0.25$. The process continues for up to 100 iterations or until the residual norm falls below $10^{-6}$.

\end{itemize}

	\begin{table}[H]
		\centering
		\caption{Relative $L_2$ errors for gravity inversion methods}
		\label{table1}
		\setlength{\tabcolsep}{9pt}
		\renewcommand{\arraystretch}{1.7}
		\begin{tabular}{llcccccc}
			\hline
			\hline
			\multicolumn{2}{c}{} & \multicolumn{3}{c}{$\boldsymbol{\displaystyle\frac{\|\rho - \hat{\rho}\|_{L_2(\Omega)}}{\|\rho\|_{L_2(\Omega)}}}$} & \multicolumn{3}{c}{$\boldsymbol{\displaystyle\frac{\|A\hat{\rho} - g\|_{L_2(\Omega)}}{\|g\|_{L_2(\Omega)}}}$} \\
			\cmidrule(lr){3-5} \cmidrule(lr){6-8}
			\multicolumn{2}{c}{} & Training & \multicolumn{2}{c}{Test} & Training & \multicolumn{2}{c}{Test} \\
			\midrule
			\multirow{1}{*}{\textbf{Direct Methods}} & \textbf{CNN} & $0.035$ & \multicolumn{2}{c}{$0.049$} & $0.0005$ & \multicolumn{2}{c}{$0.0027$}\\
			\midrule
			\multirow{2}{*}{\textbf{Generative Approaches}} 
			& \textbf{VAE} & -- & \multicolumn{2}{c}{$0.121$} & -- & \multicolumn{2}{c}{$9.2 \times 10^{-5} $} \\
			& \textbf{GAN} & -- & \multicolumn{2}{c}{$0.095$} & -- & \multicolumn{2}{c}{$3.1 \times 10^{-5} $} \\
			\midrule
			\multirow{4}{*}{\textbf{Iterative Optimization}} 
			& \textbf{GD}  &  -- & \multicolumn{2}{c}{$0.048$} & -- & \multicolumn{2}{c}{$9.8 \times 10^{-4} $} \\
			& \textbf{GMRES} &  -- & \multicolumn{2}{c}{$0.047$} & -- & \multicolumn{2}{c}{$5.1 \times 10^{-5} $} \\
			& \textbf{LGMRES}  &  -- & \multicolumn{2}{c}{$0.046$} & -- & \multicolumn{2}{c}{$3.5 \times 10^{-5} $} \\
			& \textbf{ICG}  &  -- & \multicolumn{2}{c}{$0.049$} & -- & \multicolumn{2}{c}{$4.1 \times 10^{-4} $} \\
			\hline
			\hline
		\end{tabular}
	\end{table}

\begin{table}[H]
	\centering
	\caption{Errors presented in the VAE}
	\label{table2}
	\setlength{\tabcolsep}{9pt}
	\renewcommand{\arraystretch}{1.7}
	\begin{tabular}{lccccc}
		\hline
		\hline
		\multicolumn{1}{c}{} & \multicolumn{2}{c}{ $\textbf{Reconstruction Loss}= \mathbb{E}_{q_\phi(\boldsymbol{z}|\boldsymbol{x})} [\log p_\theta(\boldsymbol{x}|\boldsymbol{z})]$} & & \multicolumn{2}{c}{$ \boldsymbol{\mathcal{D}}_{KL}\Big( q_\phi(\boldsymbol{z}|\boldsymbol{x}) \ ||\ p(\boldsymbol{z})\Big)$} \\
		\cmidrule(lr){2-3} \cmidrule(lr){5-6}
		 & Training & Test & &Training & Test \\
		 \textbf{VAE} & $110$ & $129$ & &$50$ & $51$ \\
		 \hline
		 \hline
	\end{tabular}
\end{table}

\begin{table}[H]
	\centering
	\caption{Errors presented in the GAN}
	\label{table3}
	\setlength{\tabcolsep}{9pt}
	\renewcommand{\arraystretch}{1.7}
	\begin{tabular}{lccc}
		\hline
		\hline
		\multicolumn{1}{c}{} & \multicolumn{1}{c}{$\textbf{L}_{\textbf{Generator}} = -\mathbb{E}_{\boldsymbol{z}}  [\,\log( D(G(\boldsymbol{z})))\,]$} & & \multicolumn{1}{c}{ $\textbf{L}_{\textbf{Discriminator}} = -\mathbb{E}_{\boldsymbol{x}}  [\,\log D(\boldsymbol{x})\,] - \mathbb{E}_{\boldsymbol{z}}  [\,\log(1 - D(G(\boldsymbol{z})))\,]$} \\
		\textbf{GAN} & $1.08$ &  & $0.95$  \\
		\hline
		\hline
	\end{tabular}
\end{table}

\section{Discussion}

In this paper, a range of diverse approaches for solving the gravity inversion problem has been explored. Each of these methods focuses on a different strategy; some rely purely on learning patterns from data, while others leverage mathematical optimization techniques to enhance the inversion results.
Specifically, We explored a range of methods, including direct CNN-based inversion, generative models like VAEs and GANs, and iterative techniques that enhance machine-learned estimates through numerical solvers.

By evaluating these methods under a consistent framework, our goal is to understand their relative strengths and limitations in handling the ill-posedness and complexity inherent in gravity inversion. This Section provides a case-by-case comparative analysis of the proposed approaches, using results from Tables \ref{table1}, \ref{table2} and \ref{table3} and Figures 
\ref{CNN_Gravity},
\ref{Generator_Result_Gravity},
\ref{Generator_Initial}
and
\ref{Iterative_Gravity},
to identify which techniques are best suited for specific inversion objectives such as accuracy, stability, and computational cost.

\begin{enumerate}
	\item 
	\textbf{Direct  Inversion (Deep Learning)}. This method uses a CNN to directly map 1D surface gravity measurements to 2D subsurface density fields. It demonstrates good visual and numerical performance (Figure \ref{CNN_Gravity} and Table \ref{table1}), particularly when the number of surface measurements is increased from n to 3n, which helps the network better resolve spatial details. Comparison with prior studies in this domain indicates the proposed method's significant superiority. This performance enhancement is largely attributed to utilization of a comprehensive dataset, and meticulous hyperparameter tuning of the network architecture. Furthermore, the computational cost during inference is low, making this method efficient after training. However, it does not incorporate the underlying physics of the forward model and relies entirely on the quality and diversity of the  training data. Thus, its performance is tightly coupled to how well the test samples match the learned distribution. 
	\item 
	\textbf{Generative Approaches}. In this class of methods, trained neural networks are used to constrain the solution space to a set of physically plausible and realistic density distributions.
	 After training, the generator is used as a nonlinear mapping from latent space to the physical model domain. In this framework, the inversion problem is reformulated as an optimization task in latent space. The objective becomes finding a latent vector that, when passed through the generator, yields a density model $\rho$ achieving optimal fit to observed gravity data under the forward model. In this study, two distinct generative models were trained: VAE and GAN. Figures \ref{Gen_VAE} and \ref{Gen_GAN} display random outputs from these generators, demonstrating their successful learning of the physical and real space of density distributions. Furthermore, based on Figure \ref{Generator_Power}, it can be concluded that these generators are capable of producing a wide range of physically plausible density models and have effectively captured the convex hull of the data space. 
	 However, during inversion, these generators face significant challenges. Specifically, when generating density $\rho$ to minimize $||A\rho - g||_2^2$ for given gravity data $g$, different latent vector initializations yield distinct density solutions with minimal mutual resemblance (Figure \ref{Generator_Result_Gravity}). Nevertheless, all such results maintain valid geophysical structures, satisfy essential physical constraints, and substantially reduce the error norm $||A\rho - g||_2^2$. This demonstrates severe ill-posedness of the inverse problem, where even within the physical model space, no unique solution exists. In Section \ref{generative_approaches}, multiple initialization schemes for the latent vector were implemented to assign a suitable initial value to the latent vector, yet visual results (Figure \ref{Generator_Initial}) indicate none significantly improved solution quality.

\item 
\textbf{Iterative methods with initial geuss:} This methodology investigates whether combining a suitable initial estimate (e.g., the deep learning output from Section \ref{direct_cnn}) with classical iterative methods can enhance gravity inversion performance. Although these techniques are designed based on optimization principles, their practical efficacy depends critically on both the quality of the initial estimate and the mathematical properties of the forward model. In the problem under study, the equation $A\rho = g$ represents an underdetermined system, where the dimension of the null space of matrix $A$ is significantly larger than that of its column space.
As a result, each gravity measurement corresponds to an infinite set of admissible density fields that are consistent with the linear system.
 For this reason, even when provided with a reasonably good initial guess, the visual outcomes of these algorithms exhibit minimal variation (see Figure~\ref{Iterative_Gravity}). In fact, these methods aim to minimize the numerical error by making only slight adjustments to the initial estimate, and the process terminates once a predefined level of accuracy is achieved.
 Moreover, some of these methods require squaring the system of equations (i.e., converting it to the normal equations), which may further deteriorate the conditioning of the system, as this transformation can exponentially increase the condition number.
\end{enumerate}

Overall, the evaluation of various approaches indicates that incorporating physical knowledge through current methods does not significantly enhance the performance of gravity inversion, primarily due to the inherently ill-posed nature of the problem. In contrast, purely data-driven strategies that rely on learning the statistical structure of the data can offer more effective solutions for such challenges. Further development of these approaches, as well as their integration with advanced image processing techniques, holds promise for opening new avenues in future research within this domain.

%

	\section*{Data and Code Availability}
	\noindent The datasets used in this study are publicly available at
	 \href{https://github.com/Vahid-Negahdari/Gravity_Inversion/releases}{GitHub Releases} . 
	The source code that supports the findings of this research is openly accessible at \href{https://github.com/Vahid-Negahdari/Gravity_Inversion}{GitHub}. 
	Both the dataset and code are provided for reproducibility and further research purposes.

	\bibliography{mybibfile}
	
\end{document}